\documentstyle[aps,prl,t1enc,epsf]{revtex}           %
\def \ts {\textstyle}
\def\la {\langle}
\def\ra {\rangle}
\def \D {\mbox{D}}
\def \d {\mbox{d}}
\def \t {\tilde}
\def\div {\mbox{div}\,}
\def\rd {\displaystyle{\cdot}}
\def\c {\mbox{curl}\,}
\newcommand{\ep}{\epsilon}
\newcommand{\be}{\begin{equation}}
\newcommand{\ee}{\end{equation}} 
\newcommand{\eei}{\end{equation}\indent\indent}
\newcommand{\bc}{\begin{center}}
\newcommand{\ec}{\end{center}}
\newcommand{\ber}{\begin{eqnarray}}
\newcommand{\ear}{\end{eqnarray}}
\newcommand{\ba}{\begin{array}}
\newcommand{\ea}{\end{array}}
\newcommand{\n}{\tilde{\nabla}} 
\newcommand{\na}{\nabla}
\newcommand{\hl}{\vspace{0.2cm}\hrule width\hsize height 
0.45pt\vspace{0.2cm}}
\newcommand{\hs}{\,-\,}
\newcommand{\p}{\partial}
\newcommand{\da}{{\cal D}_a}
\newcommand{\ft}{\left(\frac{t}{t_i}\right)}
\newcommand{\0}{^{(0)}}
\newcommand{\1}{^{(1)}}
\newcommand{\2}{^{(2)}}
\newcommand{\3}{{}^{(3)}}
\newcommand{\ab}{_{\alpha\beta}}
\newcommand{\h}{\case{a'}/{a}}
\newcommand{\inps}{\psi^{-1}}
\newcommand{\ua}{\left(\frac{\nabla_a\phi}{\psi}\right)}
\newcommand{\as}{{\cal A}}
\newcommand{\bb}{{\cal B}}
\newcommand{\cc}{{\cal C}}
\newcommand{\dd}{{\cal D}}
\newcommand{\kk}{{\cal K}}
\newcommand{\ls}{{\cal L}}
\newcommand{\so}{{\cal O}}
\newcommand{\sr}{{\cal R}}
\newcommand{\zz}{{\cal Z}}
\newcommand{\ti}{\tilde}
\newcommand{\wt}{\widetilde}
\newcommand{\jn}{{(j)}}
\newcommand{\inu}{ {(\nu)} }
\newcommand{\wcn}{\wt{C}_{(\nu)}}
\newcommand{\dn}{\Delta_{(\nu)}}
\newcommand{\pn}{\Phi_{(n)}}
\newcommand{\tfrac}[2]{{\textstyle{#1\over#2}}}
\def\case#1/#2{\textstyle\frac{#1}{#2} }
\newcommand{\sfrac}[2]{{\textstyle{#1\over#2}}}
\begin{document}
\draft \twocolumn[\hsize\textwidth\columnwidth\hsize\csname @twocolumnfalse\endcsname

\preprint{}

\title{Cosmological Electromagnetic Fields 
  due to Gravitational Wave Perturbations}

\author{Mattias Marklund$^{1,2,3}$\footnote{E\hs mail:
  mattias.marklund@sto.foa.se} Peter K.\ S.\ Dunsby$^2$
  \footnote{E\hs mail: peter@vishnu.mth.uct.ac.za} 
  and Gert Brodin$^4$\footnote{E\hs mail: gert.brodin@physics.umu.se}} 

\address{$1$ National Defence Research Establishment FOA, SE-172 90
  Stockholm, Sweden} 

\address{$2$ Department of Mathematics and Applied Mathematics, University
  of Cape Town, \\ Rondebosch 7701, Cape Town, South Africa}

\address{$3$ Department of Electromagnetics, 
Chalmers University of Technology, SE-412 96 Gothenburg, Sweden} 

\address{$4$ Department of Plasma Physics, 
Ume{\aa} University, SE--901 87 Ume{\aa}, Sweden} 

\date{\today}

\maketitle

\begin{abstract}
We consider the dynamics of electromagnetic fields in an 
almost\hs Friedmann\hs Robertson\hs Walker universe using 
the covariant and gauge\hs invariant approach of Ellis and 
Bruni. Focusing on the situation where deviations from the 
background model are generated by tensor perturbations only, 
we demonstrate that the coupling between gravitational waves 
and a weak magnetic test field can generate electromagnetic 
waves. We show that this coupling leads to an initial pulse 
of electromagnetic waves whose width and amplitude is determined 
by the wavelengths of the magnetic field and gravitational 
waves. A number of implications for cosmology are discussed, 
in particular we calculate an upper bound of the magnitude
of this effect using limits on the quadrapole anisotropy 
of the Cosmic Microwave Background. 
\end{abstract}

\pacs{PACS: 04.30.Nk, 95.30.Sf, 98.80.-k, 98.80.Cq}

\vskip 1pc 
]

\section{Introduction}
There have been numerous investigations on the scattering of
electromagnetic waves off gravitational fields (See Refs.\ 
\cite{bi:Collection} for a representative
sample). Most of this research have been focused on the effect 
gravitational waves have on vacuum electromagnetic fields. Other 
papers, see e.g.\ Refs.\ \cite{bi:Collection2}, also consider 
situations in Astrophysics where plasma effects are taken into
account.

Magnetic fields play an important role in our Universe, appearing
on all scales from the solar system, through interstellar and
extra\hs galactic scales, to intra\hs cluster scales of several Mpc. 
Although magnetic field inhomogeneities have not yet been observed on scales 
as large as those exhibited by Cosmic Microwave Background (CMB)
anisotropies, it is natural to expect that magnetic fields exits 
on such scales \cite{bi:Bat}, and that they could play a role in the 
formation of large\hs scale structure. Indeed many mechanisms have
been proposed to explain how these fields may be generated in the 
early universe - a process called {\it primordial magnetogenesis}. 
For example on small scales (less than the Hubble radius), QCD and 
Electro\hs Weak transitions can give rise to local charge separation 
leading to local currents which can generate magnetic fields 
\cite{bi:brustein}. Large\hs scale magnetic fields can be generated 
during inflation or in pre\hs Big Bang models based on 
string theory \cite{bi:gasperini}, in which vacuum fluctuations 
are amplified via the inflaton or dilaton.

The effect magnetic fields have on density perturbations has 
been studied extensively by a number of authors, both in 
the context of Newtonian and Relativistic Cosmology
\cite{bi:Collection3,bi:tsagas}, but as yet there have been 
no studies of their effect on gravitational wave perturbations. 

In what follows, we use the well known covariant and gauge\hs invariant 
approach of Ellis and Bruni \cite{bi:EB} to investigate this
interaction in the context of cosmology by considering the dynamics 
of electromagnetic fields in an almost Friedmann\hs Robertson\hs
Walker (FRW) universe, focusing in the situation where deviations 
from the FRW background are generated by tensor or 
gravitational wave perturbations \cite{bi:hawking,bi:DBE}. 

We show that in the presence of a weak (near\hs homogeneous) magnetic 
test field propagating on the background FRW model
\cite{bi:homogeneity}, the gravitational waves couple non\hs linearly to 
this field to produce a pulse of gravitationally induced 
electromagnetic waves. In particular, because of the different ways 
in which tensor perturbations enter the wave equations 
for the electric and magnetic fields, respectively, 
there will be, in the case of long 
wavelength gravitational waves and large\hs scale magnetic fields, 
a growth in the expansion normalised electric field, as the expansion 
normalised shear grows in time.

This paper is organised as follows. After a short discussion of 
notation and conventions, in section \ref{sec:approx} we outline 
in detail the linearisation procedure used to approximate the 
Einstein\hs Maxwell equations in cosmology. In section
\ref{sec:equations} we derive a set of non\hs linear wave equations 
which show how electromagnetic fields can be generated when
gravitational waves couple to an near\hs homogeneous magnetic test field 
propagating on the FRW background. Finally in section
\ref{sec:solutions} we solve the equations perturbatively and use 
our results to put an upper bound on the size of these 
gravitationally induced fields.   

\section{Notation and Conventions} \label{notation}
Notation and conventions are taken to be the same as in \cite{bi:Cargese}.
In particular $8\pi G=c=1$; the projected spatial covariant derivative a
tensor $T^{cd\dots}{}_{ef\dots}$ is given by $\n_a T^{cd\dots}{}_{ef\dots}
\equiv h^b{}_ah^c{}_p\dots h^d{}_qh^r{}_e\dots h^s{}_f\nabla_b 
T^{p\dots q}{}_{r\dots s}$, where $u^a$ is the 4\hs velocity of the 
matter, $g_{ab}$ is the metric tensor and $h_{ab}=g_{ab}+u_au_b$ is
the spatial projection tensor ($h_{ab}u^b=0$). A dot denotes the
covariant derivative along $u^a$, so for any tensor 
$\dot{T}^{cd\dots}{}_{ef\dots}\equiv u^a\nabla_aT^{cd\dots}{}_{ef\dots}$. 
We assume that the matter is described by {\it irrotational dust} 
\cite{bi:maartens} so that the {\it pressure} $p$, {\it acceleration
vector} $\dot{u}^a$ and {\it vorticity tensor} $\omega_{ab}$ all
vanish exactly. In this case the first covariant derivative of the 
4\hs velocity can be written as $\nabla_au_b=\sigma_{ab}+\case{1}/{3}
\Theta h_{ab}$, where $\sigma_{ab}$ and $\Theta$ are respectively the
usual shear and volume expansion of the matter congruence. 
We also define the Hubble parameter $H$ terms of the expansion 
$\Theta$ and scale factor $a$ in the usual way: $H= \Theta/3 = \dot{a}/a$. 

\section{Approximations} \label{sec:approx}
In order to simplify the non\hs linear dynamics of the coupled 
Einstein\hs Maxwell equations and to isolate the effects we are
looking for, we will adopt the following {\it approximation 
scheme} based on two parameters: $\varepsilon_{\rm g}$ will refer 
to quantities occurring in the gravitational equations, while
$\varepsilon_{\rm em}$ characterise the electromagnetic field. We 
assume that the gravitational equations follow the {\it almost FRW} 
conditions \cite{bi:EB}, so that the energy density $\mu$ and 
expansion $\Theta$ have a non\hs zero contribution in the background
model and can therefore be considered ${\cal O}(0_{\rm g})$ 
while $\sigma_{ab}$, $E_{ab}$, and $H_{ab}$ vanish in the background 
and are ${\cal O}(\varepsilon_{\rm g})$. In the case of the Maxwell 
field, we assume that there is a {\em weak magnetic test field} 
$B^a_0$ at ${\cal O}(0_{\rm em})$ which propagates on the background 
FRW model, whose gravitational influence is given by the Alfv\'{e}n 
parameter $\varepsilon \equiv (B^a_0B^0_a/\mu)^{1/2}$. On the other hand 
the electric field $E^a$ vanishes in the background and is considered to be 
${\cal O}(\varepsilon_{\rm em})$. The perturbation scheme we adopt is to drop 
terms of ${\cal O}(\varepsilon^2)$ (so that the magnetic field does not 
contribute to the gravitational dynamics \cite{bi:tsagas}), 
${\cal O}(\varepsilon_{\rm g}^2)$, ${\cal O}(\varepsilon_{\rm em}^2)$, and
${\cal O}(\varepsilon_{\rm g}\varepsilon_{\rm em})$.

In the covariant approach to linear perturbations of FRW models \cite{bi:EB}, 
{\it pure tensor} or {\it gravitational wave} perturbations are 
characterised by the following covariant conditions 
\cite{bi:hawking,bi:DBE}: 
\begin{mathletters}\label{diveqs}
\begin{eqnarray}
\n^bE_{ab}= 0~~ \Rightarrow ~~ \n_a\mu=0\,, \label{eq:divE} \\
\n^bH_{ab}= 0~~ \Rightarrow ~~ \omega_a=0\;, \label{eq:divH}
\end{eqnarray}
\end{mathletters}
the first one excludes scalar (density) perturbations and the second, 
vector (rotational) perturbations. The conditions that the terms 
on the right hand side vanish, are analogous to the transverse 
condition on tensor perturbations in the metric approach. 
In addition, we notice that since the Weyl tensor is the trace\hs free 
part of the Riemann tensor, both $E_{ab}$ and $H_{ab}$ are trace\hs free, 
again like the tensor perturbations of the Bardeen approach \cite{bardeen}. 

Given the assumed equation of state, these conditions also imply that the 
spatial gradient of the expansion $\n_a\Theta$ vanish (see \cite{bi:EB}). 
Together with (\ref{diveqs}) these conditions provide a unique 
characterisation of tensor perturbations.  

\section{Einstein\hs Maxwell Equations} \label{sec:equations}
We assume overall charge neutrality and use the Bianchi identities
and Maxwell's equations as presented in \cite{bi:Cargese}. Then, with 
the above prerequisites, we obtain a set of non\hs linear wave
equations for the gravitational ($\sigma_{ab}$: shear) and 
electromagnetic ($E^a$, $B^a$: electric and magnetic fields) 
degrees of freedom:
\begin{mathletters}\label{waveeqs}
\ber
  \Delta\sigma_{ab} + 5H\dot{\sigma}_{ab} 
    + \tfrac{3}{2}H^2\sigma_{ab} &=& 0\;, 
\label{shear} \\
  \Delta E^a + 5H\dot{E}^a 
    + 3H^2E^a 
    + \dot{j}^a + \Theta j^a&=&j^a_E
\label{E-wave} \;,\\
  \Delta B^a + 5H\dot{B}^a 
    + 3H^2B^a 
    - \epsilon^{abc}\tilde{\nabla}_bj_c 
  &=& j^a_B\;,
\label{H-wave} 
\ear
\end{mathletters}
where 
\ber
j^a_E&=& \epsilon^{abc}\tilde{\nabla}_b\left(\sigma_c\!^dB_d\right) 
+ \epsilon^{abc}\sigma^d{}_b\tilde{\nabla}_dB_c +
H^a{}_bB^b\;, \nonumber\\
j^a_B&=&2H\sigma^a{}_bB^b- 2E^a{}_bB^b+\sigma^a{}_b\dot{B}^b
\ear 
are gravitational induced magnetic and electric currents. 
and $E_{ab}=-\dot{\sigma}_{ab} - \tfrac{2}{3}\Theta\sigma_{ab}$ 
and $H^{ab}=\epsilon^{cd(a}\tilde{\nabla}_c\sigma^{b)}\!_d$ are the 
electric and magnetic parts of the Weyl tensor. Also $\Delta f \equiv 
\ddot{f}-\n^2 f$ where $f$ is any tensor orthogonal to $u^a$. 

In the above equations, the electric and magnetic fields consist
of two parts, a contribution due to the magnetic test field $B^a_0$ 
which gives rise to the current $j^a$ \cite{bi:current} and contributions 
generated by the non\hs linear coupling of this test field to 
gravitational waves via the gravitationally induced 
currents $j^a_E$ and $j^a_B$:
\ber
E^a=E^a_{\rm grav}\ , \quad B^a=B^a_0+B^a_{\rm grav}\;.
\ear

\section{Analytic Solutions and Numerical Integration}
\label{sec:solutions}
We solve the above equations perturbatively by first calculating
the gravitationally induced currents $j^a_E$ and $j^a_B$ and then
solving (\ref{E-wave}) and (\ref{H-wave}) together with (\ref{shear}) 
for the gravitationally induced electric and magnetic fields. 

To ${\cal O}(0_{\rm em})$, Maxwell's equations \cite{bi:Cargese} give 
$E=0$ and 
\be
\dot{B}_0^a+2H B^a_0=0\;,
\ee
which we can integrate to obtain:
\begin{equation}
  B^a_0 = a^{-2}A^a_{(n)} \ , \quad 
  \tilde{\nabla}^aB^b_0 = a^{-3}A^{ab}_{(n)} \ , \quad 
A^{ab}_{(n)}\equiv a\tilde{\nabla}^a A^b_{(n)}\;,
\label{eq:Aa}
\end{equation}
where $A^a_{(n)}$ and $A^{ab}_{(n)}$ determine the spatial variation 
of the magnetic test field and are constant along the fluid flow lines:
$\dot{A}^a_{(n)}=\dot{A}^{ab}_{(n)}=0$ \cite{bi:DBE}.
Furthermore we assume that the spatial functions $A^a_{(n)}$ and
$A^{ab}_{(n)}$ satisfy the Helmholtz equation
\begin{equation} \label{helmholtz2}
\tilde{\nabla}^2 A^a_{(n)} = 
-\frac{n^2}{a^2}A^a_{(n)}\;, \quad \tilde{\nabla}^2 A^{ab}_{(n)} = 
-\frac{n^2}{a^2}A^{ab}_{(n)}\;,
\end{equation}  
in this way defining a specific length scale $\lambda_{B_0}=2\pi a/n$ 
associated with the magnetic field determining its scale of inhomogeneity, 
where $n$ is a {\em fixed} wavenumber associated with that scale.

In order to solve equations (\ref{waveeqs}) it is standard to decompose 
physical (perturbed) fields into a spatial and temporal part using 
eigenfunctions which are solutions of the Helmholtz 
equation \cite{bi:harrison}. In the case of the shear tensor we write
\begin{equation}
  \sigma_{ab} = \sum_{k}\sigma_{(k)}Q_{ab}^{(k)}\;, \quad 
  \dot{Q}_{ab}^{(k)} = 0\;,  
\label{eq:shear-def}
\end{equation}
where $Q_{ab}^{(k)}$ is a tensor harmonic satisfying 
\begin{equation}\label{helmholtz1}
  \tilde{\nabla}^2Q_{ab}^{(k)} = -\frac{k^2}{a^2}Q_{ab}^{(k)}\;.
\end{equation}
We can also define higher order harmonics by taking comoving spatial
derivatives of the lower order harmonics, for example $Q_{abc}^{(k)} 
\equiv a\tilde{\nabla}_aQ_{bc}^{(k)}$ can easily be shown to satisfy
\begin{equation}\label{helmholtz3}
  \tilde{\nabla}^2Q_{abc}^{(k)} = - \frac{k^2}{a^2}Q_{abc}^{(k)}\;. 
\end{equation}

Using the above solution (\ref{eq:Aa}) (dropping the index $n$ which
indicates the scale length of $B^a_0$), 
the decomposition 
(\ref{eq:shear-def}) and writing 
\begin{equation}
  E_{\rm grav}^a = \sum_{k} {\cal E}_{(k)}{\cal E}^a_{(k)}\;, \quad 
  B_{\rm grav}^a = \sum_{k} {\cal H}_{(k)}{\cal H}^a_{(k)}\;, 
\end{equation}
the wave equations (\ref{waveeqs}) become
\begin{mathletters}\label{waveeqs2}
\be
\ddot\sigma_{(k)} + 5H\dot{\sigma}_{(k)} 
+ \left(\tfrac{3}{2}H^2  + \frac{k^2}{a^2}\right)\sigma_{(k)}=0\;,
\label{shear2}
\ee
\be
\ddot{\cal E}_{(k)} + 5H\dot{\cal E}_{(k)} + \left(3H^2
+ \frac{k^2}{a^2}+\frac{n^2}{a^2}\right){\cal E}_{(k)} 
= a^{-3}\sigma_{(k)}\;, 
\label{E-wave2}
\ee
\ber
\ddot{\cal H}_{(k)} 
&+& 5H\dot{\cal H}_{(k)} 
+ \left(3H^2 
+ \frac{k^2}{a^2}+\frac{n^2}{a^2}\right){\cal H}_{(k)}\nonumber \\ 
&=& 2a^{-2}\left(\dot{\sigma}_{(k)} 
+ 2H\sigma_{(k)}\right)\;, 
\label{H-wave2}
\ear
\end{mathletters}
where
\begin{mathletters}\label{driver}
\ber
  {\cal E}_a^{(k)}&=&
    \tfrac{3}{2}\epsilon_{abc}Q_{(k)}^{bcd}A_d
    + \tfrac{1}{2}\epsilon^{bcd}Q^{(k)}_{cda}A_b\nonumber\\
    &+& \epsilon_{abc}Q_{(k)}^{cd}A_{bd} 
    + \epsilon_{abc}Q_{(k)}^{bd}A^c{}_d \ ,
\ear
and
\be
{\cal H}_a^{(k)} = Q^{(k)}_{ab}A^b\;.  
\ee
\end{mathletters}
It is straight forward to verify that the spatial functions ${\cal E}^a_{(k)}$
and ${\cal H}^a_{(k)}$ also satisfy the Helmholtz equation.

In order to estimate the dynamical importance of our fields we introduce 
{\it expansion normalised variables} 
\begin{equation}
  \Sigma_{(k)} \equiv \frac{\sigma_{(k)}}{H} \ , \quad
  \widetilde{\cal E}_{(k)} \equiv 
    \frac{{\cal E}_{(k)}}{H} \ , \quad
  \widetilde{\cal H}_{(k)} \equiv 
    \frac{{\cal H}_{(k)}}{H} \ , \quad
\end{equation} 
giving us a set of scale invariant functions (see, e.g. \cite{bi:DynSys}). 
Introducing the conformal time parameter $\eta$ (whose defining equation is
$\dot{\eta} = a^{-1}$), the scale factor and Hubble parameter for a
dust FRW background are given by $a(\eta) = \eta^2$, $H = 2\eta^{-3}$ 
(see, e.g. \cite{bi:oldcargese}).

Substituting these into Eqs.\ (\ref{waveeqs2}) we obtain
\begin{mathletters}\label{waveeqs3} 
\be
  \Sigma_{(k)}'' + 2\eta^{-1}\Sigma_{(k)}' 
    + \left( -6\eta^{-2} + k^2 \right)\Sigma_{(k)} = 0 \ ,  
\label{shear3}
\ee
\be
  \widetilde{\cal E}_{(k)}'' 
    + 2\eta^{-1}\widetilde{\cal E}_{(k)}'
    + (k^2+n^2) \widetilde{\cal E}_{(k)} = 
\eta^{-2}\Sigma_{(k)} \ , 
\label{E-wave3} 
\ee
and
\ber
  \widetilde{\cal H}_{(k)}'' 
    &+& 2\eta^{-1}\widetilde{\cal H}_{(k)}'
    + (k^2+n^2) \widetilde{\cal H}_{(k)}
\nonumber\\ 
&=& 2\eta^{-2}(\Sigma_{(k)}' + \eta^{-1}\Sigma_{(k)}) \ .
\label{H-wave3}
\ear
\end{mathletters}

Equations (\ref{waveeqs3}) can be solved exactly in the long wave length 
gravitational wave limit (i.e.\ the term $k^2/a^2$ is small compared to 
the other terms in the wave equations), but a numerical investigation gives
more transparent results. It turns out to be convenient to introduce
the variables $A_{(k,n)} \equiv \tilde{\cal E}_{(k)}\eta_0^2/\Sigma(\eta_0)$,
$B_{(k,n)} \equiv \tilde{\cal H}_{(k)}\eta_0^2/\Sigma(\eta_0)$, where
$\eta_0$ and $\Sigma(\eta_0)$ are respectively the initial values of 
the conformal time and the normalised shear, and we have reinstated
the index $n$ indicating the scale length of $B^a_0$. These variables are 
invariant with respect to changes in $\eta_0$ and $\Sigma(\eta_0)$, 
thus giving us a scale invariant measure of the generated 
electromagnetic field. Performing the integration for various 
values of the wave number $n$ for the magnetic test field, we find 
that the normalised electric field $A_{(k,n)}$ tends to a constant 
value which depends linearly on the initial value of the shear perturbation, 
while the normalised magnetic field $B_{(k,n)}$ 
tends asymptotically to zero (see Figs. 1 \& 2 below). 

In the long wave length gravitational wave case, the expansion 
normalised shear is given by $\Sigma_{(k_{\rm long})} = \sfrac34C_1\eta^2-
3C_2\eta^{-3}$, where $C_1$ and $C_2$ are integration
constants. For late times, during the matter dominated era (when the
equation of state $p=0$ applies), the second term in $\Sigma_{(k_{\rm long})}$ 
can be neglected. In this way we can easily obtain the late time 
behaviour of the expansion normalised gravitationally induced electric field: 
\ber
\widetilde{\cal E}_{(k_{\rm long})}=\left\{
\begin{array}{c}
\frac{1}{12\pi^2}\left(\frac{\lambda_{B_0}}{\lambda_H}\right)\Sigma_{(k_{\rm
    long})}\;, n\neq 0\cr\cr 
\sfrac16\Sigma_{(k_{\rm long})}\;, n=0\;,
\end{array}
\right.
\ear
where $\lambda_H=1/H$ is the Hubble radius during that epoch.
It follows the generated electric field is proportional to the 
expansion normalised shear. Since the normalised magnetic field 
asymptotically tends to zero, the above results demonstrate 
that electric fields produced by this effect could play an important 
dynamical role in the early universe, possibly 
causing charge separation. Furthermore, because the 
asymptotic value of the electric field is proportional to 
the magnitude of the shear, we can use the CMB anisotropy limits 
on $\Sigma$ to give an upper bound on the size of this effect 
\cite{bi:SAG}:
\ber
\widetilde{\cal E}_{(k_{\rm long})}\lesssim\left\{

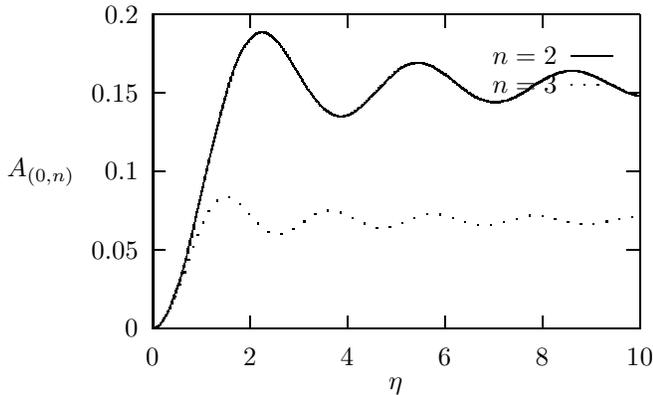
  \caption{Expansion normalised gravitationally induced electric field 
    for different values of the magnetic wavenumber $n$.}
\end{figure}

\begin{figure}
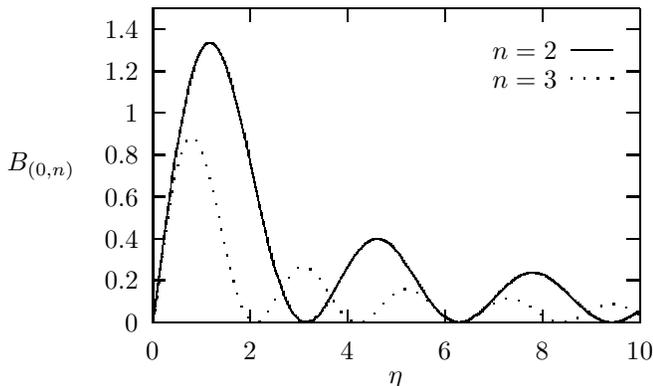

\centering
\setlength{\unitlength}{0.240900pt}
\ifx\plotpoint\undefined\newsavebox{\plotpoint}\fi
\sbox{\plotpoint}{\rule[-0.200pt]{0.400pt}{0.400pt}}%
%
  \caption{Expansion normalised gravitationally induced magnetic field 
for different values of the magnetic wavenumber $n$.}
\end{figure}

\section{Discussion}
In this paper we derived a set of non\hs linear wave equations 
which demonstrate how electromagnetic fields can be generated when
gravitational waves couple to an near\hs homogeneous magnetic test field 
propagating on a FRW background. In particular we found that for 
long wavelength gravitational waves, the gravitationally induced
fields are proportional to the magnitude of the expansion normalised 
shear which characterise tensor perturbations. This allows a simple 
determination of an upper bound on the magnitude of these fields 
based on the quadrapole anisotropy of the Cosmic Microwave Background.

We note that this paper has not considered the back\hs reaction of
this effect on the gravitational dynamics, which although small may 
also give rise to interesting results. This issue will be  
explored in a forthcoming paper \cite{bi:DMB}.

\section*{Acknowledgements}

We thank Roy Maartens and the Referee for useful comments.
M.\ M.\ was supported by the Royal Swedish Academy of Sciences. P.\
K.\ S.\ D.\ was supported by the NRC (South Africa).




\begin{references}
\bibitem{bi:homogeneity} Although this 
in principle breaks the homogeneity and isotropy of the FRW 
spacetime, a week magnetic test field can be shown to be consistent 
in such a background provided the the spatial variations it generates 
are negligible.
\bibitem{bi:current} Note that the inhomogeneity variable 
$\tilde{\nabla}_aB_b$ is determined by the background current $j^a$.
\end{references}
\end{document}